# GROWTH OF QUANTUM WELL FILMS OF TOPOLOGICAL INSULATOR BI₂SE₃ ON INSULATING SUBSTRATE

CUI-ZU CHANG, KE HE[*], LI-LI WANG AND XU-CUN MA

*Institute of Physics, Chinese Academy of Sciences, Beijing 100190, China*
*[*]kehe@aphy.iphy.ac.cn*

MIN-HAO LIU, ZUO-CHENG ZHANG, XI CHEN, YA-YU WANG, AND QI-KUN XUE[*]

*Department of Physics, Tsinghua University, Beijing 100084, China*
*[*]qkxue@mail.tsinghua.edu.cn*

Insulating substrates are crucial for electrical transport study and room temperature application of topological insulator films at thickness of only several nanometers. High quality quantum well films of $Bi_2Se_3$, a typical three-dimensional topological insulator, have been grown on $\alpha$-$Al_2O_3$ (sapphire) (0001) by molecular beam epitaxy. The films exhibit well-defined quantum well states and surface states, suggesting the uniform thickness over macroscopic area. The $Bi_2Se_3$ thin films on sapphire (0001) provide a good system to study low-dimensional physics of topological insulators since conduction contribution from the substrate is negligibly small.

*Keywords*: A list of 3–5 keywords are to be supplied.

## 1. Introduction

Recently, topological insulators (TIs) have attracted much attention in condensed matter physics and material sciences due to their theoretically predicted exotic physical properties and possible applications in spintronics and quantum computation.[1-13] TI films with a thickness of only several nanometers are especially interesting for they are building blocks for TI-based heterostructures and planar devices, and could exhibit novel properties, e.g. enhanced thermoelectric[14], quantum anomalous Hall effect[15], and exciton condensation[16]. $Bi_2Se_3$, $Bi_2Te_3$, and $Sb_2Te_3$ are among the most studied three-dimensional TIs for their simple surface band structure and relatively large bulk gap, which makes them possibly work even at room temperature.[9,10] High quality quantum films of $Bi_2Se_3$ family TIs with controlled thickness have been realized with molecular beam epitaxy (MBE) techniques.[17-19] So far, the substrates used for the MBE growth of $Bi_2Se_3$ family such as Si (111) and SiC (0001) are all semiconducting.[17-19] For the possible room temperature applications of thin films of $Bi_2Se_3$ TIs, semiconducting substrates are not satisfactory since they will contribute considerable conductivity. Even for the electrical transport studies at low temperature, the conductivity contribution from the buffer layer and/or the space charge layer at the interface cannot be neglected, can easily override that from the topological surface states.

In this work, we report layer-by-layer MBE growth of $Bi_2Se_3$ thin films on $\alpha$-$Al_2O_3$ (sapphire) (0001) substrate. Well-defined quantum well states and surface states are revealed by *in situ* angle-resolved photoemission spectroscopy (ARPES), demonstrating that the $Bi_2Se_3$ films grown on sapphire (0001) have as good quality as those on graphene-terminated SiC (0001).[9,10] Thanks to the excellent insulating property of the substrates, reliable electrical transport measurement on the $Bi_2Se_3$ thin films is realized.[20]

## 2. Experimental

$Bi_2Se_3$ thin films were grown in an ultra-high vacuum (UHV) system (<1.5×10⁻¹⁰ torr) consisting of a MBE chamber, an Omicron STM, and an ARPES system. ARPES was carried out with a Gamma-data ultraviolet lamp and a Scienta SES-2002 analyzer. The substrates used for $Bi_2Se_3$ growth are commercial sapphire (0001) (Shinkosha Co., Japan). Prior to sample growth, sapphire substrates were first degassed at 650℃ for 90 minutes and then heated at 850 °C for 15 minutes. High purity Bi (99.9999%) and Se (99.999%) were evaporated from standard Knudsen cells. To avoid charging of the samples caused by the insulating substrate in ARPES and STM measurements, a 300-nm-thick titanium layer was deposited both edges of the substrate, which is connected to the sample holder. Once a continuous film is formed it is grounded through the contacts. All STM and ARPES measurements were carried out at room temperature. The transport measurements were taken in a cryostat with magnetic field B up to 15 Tesla and sample temperature down to 1.5K.

## 3. Results and Discussion

Figure 1(a) shows the reflective high energy electron diffraction (RHEED) pattern of the sapphire substrate. Sharp 1×1 diffraction streaks and clear Kikuchi lines demonstrate the high crystalline quality of the sapphire surface. Similar to MBE growth of $Bi_2Se_3$ films on graphene-terminated SiC (0001), [17,18] a Se-rich growth condition was used (Bi/Se flux ratio is about 1:15) to obtain high quality stoichiometric films with the substrate temperature set between that of Se source and that of Bi source. Figure 1(b) shows the RHEED pattern of a $Bi_2Se_3$ film of 25 quintuple layers (QLs). Again, a clear and sharp 1×1 pattern appears. The [1 0 -1 0] axes of the $Bi_2Se_3$ film and sapphire substrate are parallel to each other. We can see that in spite of the large lattice mismatch (~13%), single crystal $Bi_2Se_3$ film can still be obtained. It is probably due to the layered structure of $Bi_2Se_3$, which easily relaxes the strain induced by interface mismatch. The

ARPES band map (second order differential spectra) of the film is shown in Fig. 1(c). Dirac cone centered at Γ point can be clearly observed. The Dirac point locates at 135 meV below the Fermi level, very close to the doping level of the $Bi_2Se_3$ films grown on graphene-terminated SiC (0001), indicating the high quality of the film.[17]

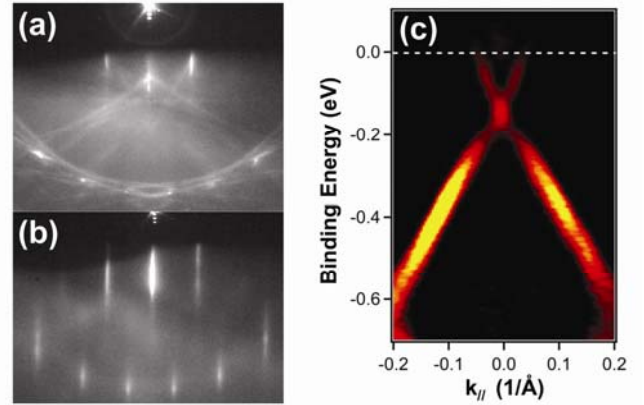

Fig. 1. RHEED patterns of (a) sapphire (0001) and (b) 25QL $Bi_2Se_3$ film grown on sapphire (0001). (c) ARPES spectra of a 25QL $Bi_2Se_3$ film on sapphire (0001).

In order to obtain $Bi_2Se_3$ quantum well films, we optimized the growth condition further with STM. Figure 2 exhibits a series of STM images of nominal 8QL $Bi_2Se_3$ films grown at the same Bi/Se flux ratio but different substrate temperatures ($T_{sub}$). At $T_{sub}$ =190 °C [see Fig. 2(a)], STM image shows elongated islands with a height of 3~4 nanometers. Besides, we can see atomic flat terraces with basically only 1 QL height difference at most. Similar elongated islands can also be observed on MBE grown Bi films on graphite substrate.[21] So they are probably formed by additional Bi atoms on the surface. Although the flux of Se is much higher than that of Bi, the low substrate temperature can lead to inefficiency in the dissociation of Se molecules and the reaction between Bi and Se atoms. Actually the elongated islands disappear at higher substrate temperature [see Figs. 2(b) and 2(c)]. But when the substrate temperature is too high [see the STM image for the $T_{sub}$ = 250°C sample shown in Fig. 2(c)], a lot of screw islands appear. High density of screw dislocation is not strange in MBE growth of the large lattice mismatched system such as GaN/sapphire. By lowering the substrate temperature into non-equilibrium condition, the density of screw dislocation could be reduced. Here in a small temperature window around 220$^o$C [see Fig. 2(b)], both the elongated Bi islands and the screw dislocations are nearly removed from the surface of Bi$_2$Se$_3$ film.

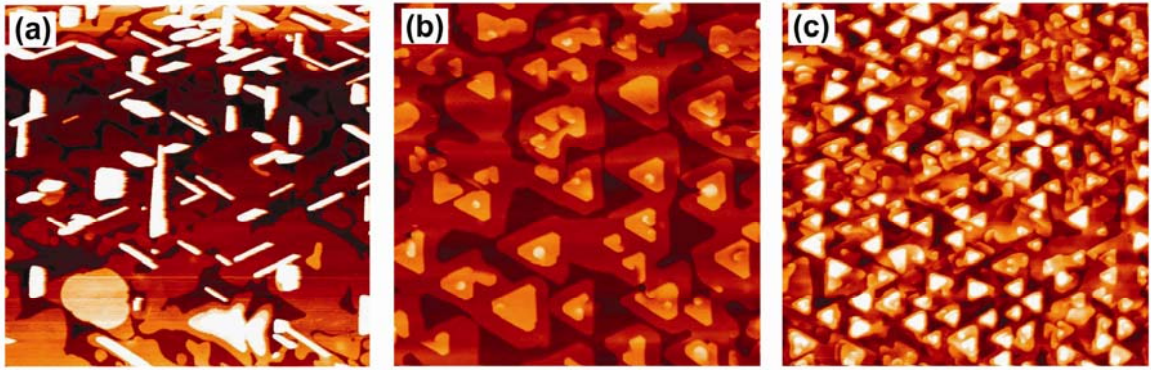

Fig. 2. STM morphology (1μm×1μm) of nominal 8QL Bi$_2$Se$_3$ film grown on sapphire (0001) with substrate temperature of (a) 190$^o$C, (b) 220$^o$C, and (c) 250$^o$C, respectively.

At the optimized substrate temperature, layer-by-layer growth of Bi$_2$Se$_3$ film is realized. Figures 3(a)-3(f) show a series of ARPES band maps of Bi$_2$Se$_3$ films with thickness from 1 to 6 QL. All spectra are taken along the Γ-K direction. The ARPES data are very similar to that of the films on graphene-terminated SiC (0001) substrate.[17,18] In the spectra for 1QL film, only a parabolic electron band can be observed. At 2QL, the spectra show a hole band and an electron band spaced by a gap of 245 meV. With increasing thickness, the hole band and the electron band gradually approach to each other, and finally merge into a Dirac cone at 6QL. The gap-opening in the Dirac surface states below 6QL is the result of the coupling between the surface states from the surface and interface sides of the film[17].

Quantum well states of both conduction band and valence band can be clearly seen above and below the surface bands, respectively, suggesting the uniform thickness of the films. Rashba-type splitting, which has been observed in the Bi$_2$Se$_3$ films from 3QL to 5QL on graphene-terminated SiC (0001) substrate[17], cannot be resolved here. The Rashba splitting results from the different chemical potentials between the interface and surface of the film induced by the substrate. Its absence suggests that the Bi$_2$Se$_3$ films grown on sapphire (0001) have less asymmetry, which might be attributed to the rather large energy gap of Al$_2$O$_3$.

Insulating substrate is very important for obtaining correct transport results from topological insulator thin films. Although Bi$_2$Se$_3$ films grown on graphene-terminated SiC (0001) have been shown to have very high crystalline quality, the existence of graphene layer at the interface makes the interpretation of transport results difficult. Figure 4 shows the temperature dependences of resistance of a graphene-terminated SiC (0001), a 10 nm thick Bi$_2$Se$_3$ film grown on graphene-terminated SiC (0001), and a 10 nm thick Bi$_2$Se$_3$ film grown on sapphire (0001). We can see that at low temperature the resistance of the graphene-terminated SiC (0001) is at the same order with that of the Bi$_2$Se$_3$ film on sapphire (0001). Since the SiC (0001) substrate used here has a very low doping level (the resistivity ~10$^6$ Ω·cm), the high conductivity can only be attributed to the graphene layer. Actually the conductivity of the Bi$_2$Se$_3$ film on graphene-terminated SiC (0001) is roughly the sum of that of the graphene-terminated SiC (0001) and that of the Bi$_2$Se$_3$ film on sapphire (0001). Hall effect measurement at 1.5 K [see inset of Fig. (4)] also reveals much higher carrier density in the Bi$_2$Se$_3$ film on graphene-terminated SiC (0001) than that in the Bi$_2$Se$_3$ film of the same thickness on sapphire (0001), which must be attributed to graphene layer. Therefore, for transport measurement of Bi$_2$Se$_3$ thin films of only several QLs, sapphire (0001) is a much better substrate than the graphene-terminated SiC (0001) for its negligible conduction contribution.

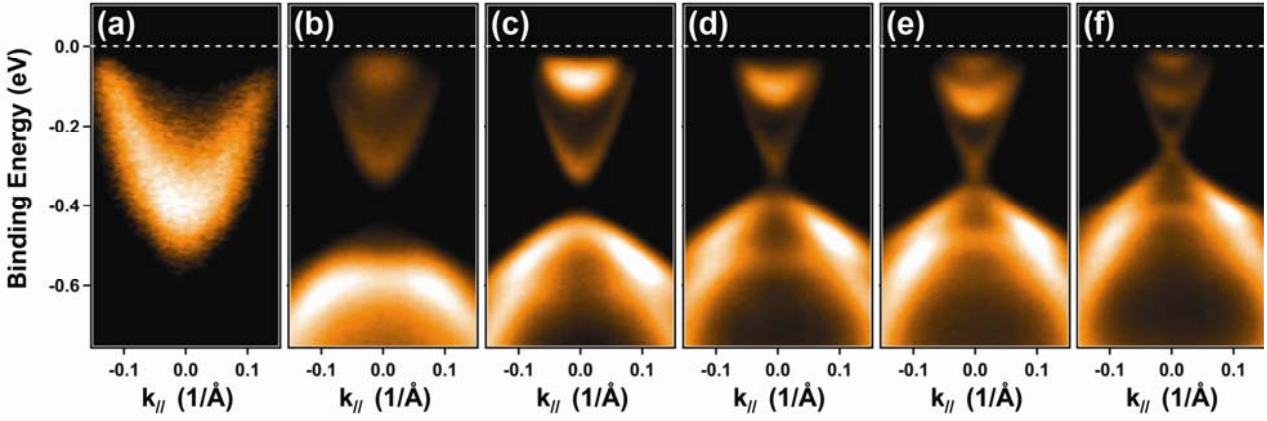

Fig. 3. ARPES spectra of (a) 1QL, (b) 2QL, (c) 3QL, (d) 4QL, (e) 5QL, and (f) 6 QL thick Bi$_2$Se$_3$ films grown on sapphire (0001). All the spectra are taken along Γ-K direction. The dashed line indicates the Fermi level.

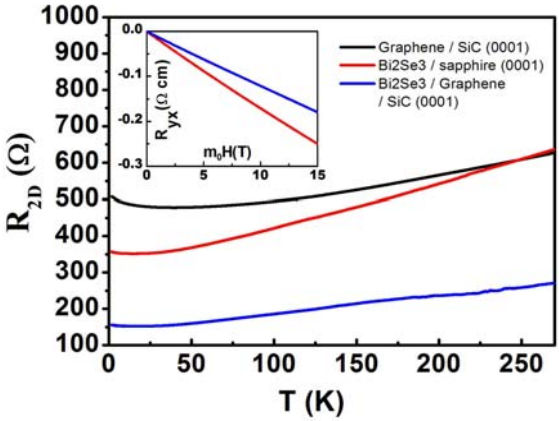

Fig. 4. Temperature dependence resistance of a graphene-terminated SiC (0001) (black), a 10 QL Bi$_2$Se$_3$ film grown on graphene-terminated SiC (0001) (red), and a 10 QL Bi$_2$Se$_3$ film grown on sapphire (0001) (blue). The inset shows the result of Hall measurement of 10 QL thick Bi$_2$Se$_3$ films on graphene-terminated SiC (0001) (blue) and sapphire (0001) (red), respectively.

Growth of Bi$_2$Se$_3$ films on other insulating substrates like SrTiO$_3$ (111) and mica (0001) has also been tried. Dirac cone-shaped surface states were observed in the films on both substrates by ARPES. But the surface state peaks are much broader than those on sapphire (0001), indicating the lower crystalline quality. It may result from the impurity atoms diffusing from the substrates since both substrates contain several elements. Nevertheless two substrates are good as back gate in tuning the carrier density of TIs, which could be useful in future TI-based devices.

## 4. Summary

We have shown that sapphire (0001) is a good insulating substrate for growth of high quality Bi$_2$Se$_3$ films in quantum well regime. The Bi$_2$Se$_3$ films grown sapphire (0001) serve as a good system to study the transport properties of topological insulators [20] and provide a basis structure of TI-based devices.

### Acknowledgments

This work was supported by NSFC and MOST of China.

### References

1. X. L. Qi, and S. C. Zhang, Phys. Today **63**, 33 (2010).
2. J. E. Moore, Nature **464**, 194 (2010).
3. X. L. Qi, and S. C. Zhang, arXiv: 1008.2026 (2010).
4. M. Z. Hasan and C. L. Kane, Rev. Mod. Phys. **82**, 3045 (2010).
5. B. A. Bernevig, T. L. Hughes, and S. C. Zhang, Science **314**, 1757(2006).
6. L. M. König, S. Wiedmann, C. Brüne, A. Roth, H. Buhmann, L. W. Molenkamp, X. L. Qi, and S. H. Zhang, Science **318**, 766 (2007).
7. D. Hsieh, D. Qian, L. Wray, Y. Xia, Y. S. Hor, R. J. Cava, and M. Z. Hasan, Nature (London) **452**, 970 (2008).
8. D. Hsieh, Y. Xia, L. Wray, D. Qian, A. Pal, J. H. Dil, J. Osterwalder, F. Meier, G. Bihlmayer, C. L. Kane, Y. S. Hor, R. J. Cava, and M. Z. Hasan, Science **323**, 919 (2009).
9. H. J. Zhang, C. X. Liu, X. L. Qi, X. Dai, Z. Fang, and S. C. Zhang, Nat. Phys. **5**, 438 (2009).
10. Y. Xia, D. Qian. D. Hsieh, L. Wray, A. Pal, H. Lin, A. Bansil, D. Grauer, Y. S. Hor, R. J. Cava, and M. Z. Hasan, Nat. Phys. **5**, 398 (2009).

11. D. Hsieh, Y. Xia, D. Qian, L. Wray, J. H. Dil, F. Meier, J. Osterwalder, L. Patthey, J. G. Checkelsky, N. P. Ong, A. V. Fedorov, H, Lin, A. Bansil, D. Grauer, Y. S. Hor, R. J. Cava, and M. Z. Hasan, Nature (London) **460**, 1101 (2009).

12. Y. L. Chen, J. G. Analytis, J. H. Chu, Z. K. Liu, S. K. Mo, X. L. Qi, H. J.Zhang, D. H. Lu, X. Dai, Z. Fang, S. C. Zhang, I. R. Fisher, Z. Hussain and Z. X. Shen, Science **325**, 178 (2009)

13. L. Fu, C. L. Kane, and E. J. Mele. Phys. Rev. Lett. **98**, 106803 (2007).

14. Pouyan Ghaemi, Roger S. K. Mong, and J. E. Moore. Phys. Rev. Lett. **105**, 166603 (2010).

15. R. Yu, W. Zhang, H. J. Zhang, S.C. Zhang, X. Dai, Z. Fang. Science **329**, 61 (2010).

16. B. Seradjeh, J. E. Moore, and M. Franz. Phys. Rev. Lett. **103**, 066402 (2009).

17. Y. Zhang, K. He, C. Z. Chang, C. L. Song, L. L. Wang, X. Chen, and J. F. Jia, Z. Fang, X. Dai, W. Y. Shan, S. Q. Shen, Q. Niu, X. L. Qi, S. C.Zhang, X. C. Ma, and Q. K. Xue, Nat. Phys. **6**, 584 (2010).

18. C. L. Song, Y. L. Wang, Y. P. Jiang, Y. Zhang, C. Z. Chang, L. L. Wang, K. He, X. Chen, J. F. Jia, Y. Y. Wang, Z. Fang, X. Dai, X. C. Xie, X. L. Qi, S. C. Zhang, Q. K. Xue, and X. C. Ma, Appl. Phys. Lett. **97**, 143118(2010)

19. Y. Y. Li, G. Wang, X. G. Zhu, M. H. Liu, C. Ye, X. Chen, Y. Y. Wang, K. He, L. L. Wang, X. C. Ma, H. J. Zhang, X. Dai, Z. Fang, X. C. Xie, Y. Liu, X. Q. Qi, J. F. Jia, S. C. Zhang, and Q. K. Xue, Adv. Mater. **22**, 4002(2010).

20. M. H. Liu, C. Z. Chang, Z. C. Zhang, Y. Zhang, W. Ruan, K. He, L. L. Wang, X. Chen, J. F. Jia, S. C. Zhang, Q. K. Xue, X. C. Ma, and Y.Y. Wang, arXiv: 1011.1055 (2010).

21. S. A. Scott, M. V. Kral, and S. A. Brown, Appl. Surf. Sci. **252**, 5563 (2006).